\documentclass{article}
 \usepackage{amssymb}

 \usepackage{amscd}
\usepackage{makeidx}
\usepackage{graphicx}
\usepackage{amsmath}
\usepackage{epsfig}
\usepackage[dvips]{color}

\makeatletter
 \@addtoreset{equation}{section}
 \makeatother

\def\varepsilon{{\Greekmath 0122}}
\def\FindBoldGroup{{\setbox0=\hbox{$\mathbf{x\global\edef\theboldgroup{\the\mathgroup}}$}}}
\def\Greekmath#1#2#3#4{\if@compatibility
        \ifnum\mathgroup=\symbold
           \mathchoice{\mbox{\boldmath$\displaystyle\mathchar"#1#2#3#4$}}
          {\mbox{\boldmath$\textstyle\mathchar"#1#2#3#4$}}
      {\mbox{\boldmath$\scriptstyle\mathchar"#1#2#3#4$}}
    {\mbox{\boldmath$\scriptscriptstyle\mathchar"#1#2#3#4$}}        \else
           \mathchar"#1#2#3#4        \fi
    \else
        \FindBoldGroup
        \ifnum\mathgroup=\theboldgroup
\mathchoice{\mbox{\boldmath$\displaystyle\mathchar"#1#2#3#4$}}
       {\mbox{\boldmath$\textstyle\mathchar"#1#2#3#4$}}
{\mbox{\boldmath$\scriptstyle\mathchar"#1#2#3#4$}}
{\mbox{\boldmath$\scriptscriptstyle\mathchar"#1#2#3#4$}}
\else
           \mathchar"#1#2#3#4        \fi
      \fi}
\newif\ifGreekBold\GreekBoldfalse\let\SAVEPBF=\pbf
\def\pbf{\GreekBoldtrue\SAVEPBF}
\numberwithin{equation}{section}

 \newcount\timehh\newcount\timemm
 \DeclareMathOperator{\R}{\mathbb{R}}
 \DeclareMathOperator{\C}{\mathbb{C}}

\newcommand{\bs}{\begin{subequations}}
\newcommand{\es}{\end{subequations}}

\begin{document}
\baselineskip0.50 truecm  \font\sm =cmbx12 at 10pt
   \def\a{\alpha}

\def\b{\bar}\def\be{\beta}
\def\B {\Bbb}
 \def\ca{\cap}\def\capp{\un{n=1}\to{\ov{\inf}\to\cap}}

\def\cu{\cup}  \def\cupp{\un{n=1}\to{\ov{\inf}\to\cup}}
\def \C{\mathcal }
\def\d{\delta}
\def\D{\Delta}
\def\do{\downarrow}
\def\e{\epsilon}
\def\em{\emptyset}
\def\eq{\equiv}
\def\ex{\exists}
\def\f{\frac}\def\fo{\forall}
\def\g{\gamma}\def\G{\Gamma}\def\h{\hskip2truecm}
\def\i{\int}
\def\fty{\infty  }
\def\l{l}
\def\lm {\un{n \to\infty}\to\lim}

\def\lb{\lbrace} \def\L{\Leftarrow}
 \def\Lr{\Leftrightarrow}
\def\ls{\un{n=1}\to{\ov{\inf}\to\cap}\un{j=n}\to{\ov{\inf}\to\cup}}
\def\li{\un{n=1}\to{\ov{\inf}\to\cup}\un{j=n}\to{\ov{\inf}\to\cap}}
\def\m{\mu}
\def\n{\nu}

 \def\o{\omega}\def\ov{\overset}
\def\p{\partial}

\def\r{\rightarrow}\def\rb{\rbrace}\def\R{\Rightarrow}
\def\s{\sigma}\def\S{\Sigma}

\def\un{\underset}\def\Un{\un{n=1}\to{\ov{\inf}\to\cap}}
  \def\up{\uparrow}

\def\z{\zeta}
 \def \ta{Sean  $X_1 ,  ...,  X_n $ una coleccion de
variables aleatorias independientes y equidistribuidas}\def
\tb{Sean $X_1,  ...,  X_n, $ una co leccion de  variables
aleatorias independientes}
\def\z{\zeta}
\def\centerline#1{\line{\hss#1\hss}}
 \def\limxt0{\displaystyle\lim_{x \to 0}}
 \def\prf{$\underline{\hbox{\sl Proof}}$}
 \def\sg{\mathop{\rm ~~sign}}
 \def\Rl{\mathop{\rm ~~R}\!\!\!\!\vrule height2.1pt width2pt
depth1.2pt \, \, }
 \def\bx{\tilde x}
 \def\bxp{{\tilde x}'}
 \def\bxpp{{\bar x}''}
 \def\sxy{\sigma x-y}

 \parindent=0pt
\hfuzz=2pt
\def\hline{\noalign{\hrule}}
\let\\\par
\def\emph#1{{\It #1}}
\def\with{\small (with }
\def\({\/{\rm ~~(}}
\def\){\/{\rm ~~)}}
\def\strut{\vrule width0.5pt height 3.5pt depth 2.5pt}

\noindent {\bf    Stochastic model for market stocks with strong resistances }\\


\vskip2truecm

\noindent{\small    \underline{Javier  Villarroel}   }\\ \\


\noindent{\scriptsize   Univ.  de Salamanca, Fac. de  Ciencias, \\
Plaza Merced s/n,  Salamanca 37008, Spain \\
(Javier@usal.es)\\

}


\medskip

{\bf  PACS}

  0.7.05 Mh,     89.65.Gh, 02.50.Ey,  05.40.Jc,  \vskip0.5truecm

   \textbf{  Abstract. } {\bf  }

\noindent We   present  a  model  to describe  the  stochastic
evolution of  stocks that show a strong
  resistance  at some level and generalize to this situation the   evolution  based
upon geometric Brownian motion. If volatility and drift are
  related in a certain way we show that our model  can be integrated in an exact
way.  The related  problem of how to prize general    securities
that
 pay  dividends at a continuous  rate and  earn a terminal payoff   at maturity $T$
is solved via the martingale probability approach.

\vskip1.5truecm

 \section{\bf   Introduction}    We   consider an ideal model of  financial market
consisting of two securities: a savings account $Z_t$
 evolving via $dZ_t=r_tZ_t dt$, where
      $r_t$  is the   instantaneous interest rate of the market and     is assumed
to be deterministic (but not necessarily constant);   and   a
 "risky" asset whose price at time $t$:  $ X_t$,    evolves according to some
stochastic differential eq. (SDE) driven by Brownian motion (BM).
As
 it is well known, the prototype model for stocks-price evolution
  assumes that the return process  $R_t=\log X_t$ follows a  random walk or BM
with drift and hence that prices $X_t$ evolve via   the popular
      geometric Brownian motion (GBM) model, i.e., that    $X_t$
 satisfies
 $$dX_t=\mu X_tdt+\s X_t dW_t\eqno(1)$$
  Here $ \mu $ is the  mean return rate and $\s$ the volatility which are supposed
to be constants  while $W_t$ is a Brownian motion under the
empirical or
 real world probability. We remark that here and elsewhere in this article
integrals and SDE's are understood in the sense of It\^{o}'s
calculus.
 Transition to standard (Stratonovitch) calculus  can be done if wished.

 The solution to this SDE is given by
 $$X_t^{\text{GBM}}= x_0\exp\Big\{\s W_t+ \left(\mu-
 \f{\sigma^2 }2\right)   t \Big\}  \eqno(2) $$

\medskip

  After the seminal  work of Black and Scholes [1]
and Merton [2], who    derive a formula to price options on stocks
with underlying dynamics based upon GMB,  eq.  (1)   has   become
the paradigmatic model to describe both price evolution and
derivatives pricing.
  However,  while  such a simple model captures well the basic features of prices it
does not quite account  for more
 stylized facts that empirical prices show; among them we mention
 the  appearance  of "heavy tails" for long values  of  the relevant  density
probability distributions of returns [3,4]; further, the empirical
distribution  shows an exponential form for moderate values of the
returns,  which is not quite fitted by  the predicted log-normal
density implied by (2). The existence of
   self-scaling and long memory effects was first noticed in [5].  Due to all this
option pricing under this GBM framework can not fully account for the observed market option prices and the classical  Black-Scholes $\&$ Merton
(BSM) formula is found to overprice (respectively underprice) "in (respectively, out of) the money options".
 Apparently, for empirical prices of call options to fit this
formula  an extra dependence in the strike price, the volatility
smile, must be introduced by hand.

After the seminal paper  by   Mantegna  and  Stanley who studied
the empirical evolution  of the stock index S$\&$P500 of the
American Stock Exchange during a five year period,    several
authors have elaborated on  the possibility that price dynamics
   involves
   Levy process   and have discussed  option pricing in  such a framework (See
[5-13]). For  complete accounts of    option pricing
    and  stochastic calculus from the economist and, respectively,  physicist,
points of view see
  [14-17] and [18-21].

\medskip

  Here we shall focus  in another different aspect   that some traded stocks seem to
present, viz the possibility  of having, at some level, strong
  resistances
both  from above or below.   For example, corporations or major
institutions may have laid out  a policy under which heavy  buy
orders are triggered whenever  the stock price hits this level.
Such feature can not be described with  Eq. (1) as under such an
evolution prices can reach any value in $(0, \infty)$. Concretely,
in this paper we want to model the evolution of a market stock
which has a strong lower resistance at some level $c$ where we
suppose that $c$ is a constant.

\medskip

In section (2) we present  a model that incorporates an \it
attainable \rm  barrier at the point $x=c>0$ and hence can, in
principle, be used to account for such a fact.  We next derive the
evolution of the asset and the probability distribution function.
It turns out that $c$ is a regular barrier in terms of  Feller's
boundary theory and hence a prescription on how to proceed once
reached must be given.
       In section (3) we study pricing of securities under such a  model and obtain
a closed formula for valuation of European derivatives that
       have, in addition, a continuous stream of payments. We   tackle this problem
using
     the Martingale formalism of
       Harrison et al [22] and obtain  the partial differential equation (PDE) that
the price of a security satisfies. Solving this eq. corresponding
to particular
       final conditions we obtain the price of options under this model. This price
is compared  with that given by   the standard Black$\&$ Scholes-
Merton
       formula.   In the  appendix we consider some technical issues concerning
value of the market price of risk  and the the existence
       of the martingale measure or risk free probability under which securities are
priced.

    \medskip

\medskip

\section{Price evolution under the martingale probability}

 Let $r_t$   be  the  deterministic  interest rate at time $t$ and $Z_t=\exp
\int_0^tr_sds$   be a "savings account  ". As we pointed out we
  consider  that   $ X_t$ is the $t$-price of a
  tradeable  asset   that has a strong lower resistance at some  constant level $c$
where $0<c<x_0\eq X_0$.  Mathematically this implies that the
values  of $X_t$
  must be  restricted to the interval $[c,\infty)$
 and hence $X_t$ must have  a boundary point
 of a certain kind at $x=c$. From intuitive financial arguments the boundary can not
be of absorbing type since in that case, once
  reached, the price $X_t$ remains there.
  Further  it seems reasonable to assume that there exists positive probability to
attain the boundary; we suppose that this event  "triggers" bid
orders and hence
  that $X_t$ ricochets  upon  hitting the boundary.    Therefore in such situation
 the assumption that prices evolve via Eq. (1)  is no longer
  valid. The
obvious  modification  wherein    prices evolve as
$$ \tilde X_t^{\text{GBM}}\eq c+(x_0-c) e^{\s W_t+ \left(\mu-
 \f{\sigma^2 }2\right)   t }\eqno (3)$$ is  also  ruled out as this evolution implies that
    $c<X_t<\infty$ but the value $X_t=c $ is never attained and the probability to
get arbitrarily close to the barrier tends to zero with the
distance to
    it. Thus trajectories corresponding to a such model never quite seem to reach the support(In terms of Feller's theory briefly reminded below $x=c$ is a natural
    barrier  at which   Feller functions blow up).

    \medskip

         Motivated by similar ideas in the context of the  Cox-Ingersoll-Ross model
of interest rate dynamics [23] we now introduce a   more
satisfactory model
         which satisfies  the aforementioned features    and is at the same time
analytically
         tractable; we shall suppose that $X_t$ evolves via   the SDE

 $$dX_t=    a(t,X_t)    dt+    b(t, X_t)   d    W_t,\text{ where }    a(t,x)=\mu    x, ~~  b(t, x)=\s(t)\sqrt{     x^2-    c^2   }\eqno (4)$$
       Here  $X_0=x_0 >c$, $\mu$ is the stock
mean rate of return  and  $ b(t, x)$ the  volatility   coefficient.
    Indeed, under such  a dynamics it follows from (4) that as $x$ approaches the
point $c$, $ b(t, x)$
    tends to zero and hence $X_t$ evolves roughly like $dX_t=     \mu   X_t
    dt$ implying that $X_t$ will increase and then  escape  from the
    boundary.

 For valuation purposes
     one needs to consider the evolution  under a new probability that might be
different to the empirical observed
    probability.  Mathematically speaking a such a probability is defined requiring
that under it the discounted prices $X'_t\eq X_t/Z_t$ are
martingales (this
    {\it   "risk-neutral" probability } was introduced in [22] although the
underlying idea  pervades  the original work of  Black-Scholes
$\&$ Merton
    [1,2]).  Stated another
way,
    this means that {\it under the "risk-neutral" probability, the stock
    price $X_t$
evolves, on average, as the riskless security $Z_t$  thereby
preventing  arbitrage opportunities}. Indeed,    the martingale
property implies
 $$    e^{-\int_0^t r_sds}\Bbb  E^*  \Big(   X_t \Big|X_0 \Big)= \Bbb  E^*  \Big(
X'_t \Big|X_0
 \Big)=\Bbb  E^*  \Big(   X'_t \Big|X'_0
 \Big)=X'_0=x_0 \eqno(5)$$
where $\Bbb  E^*  \Big(   X_t \Big|X_0 \Big)$ is the conditional
average of $X_t$ given $X_0$ with respect to the martingale
probability. Hence $$    \Bbb  E^*  \Big(   X_t \Big|X_0
\Big)=x_0Z_t \eqno(6)$$

More generally, given the past history  $\mathcal F_s$ of the
process up to time $s$ (i.e., the $\s$-field of past events) one
has
$$    \Bbb  E^*  \Big(   X_t \Big| \mathcal F_s
\Big)  =  Z_t\Bbb  E^*  \Big(   X'_t \Big| \mathcal F_s
 \Big)= Z_t X'_s  \eq X_s  e^{\int_s^t r_ldl}\eqno (7) $$
   We shall assume that our market is efficient, i.e., that the martingale
probability  $   \Bbb   P^* $ exists-- which    is not always the
case.
   In such a case the explicit form of the original  drift
   coefficient
  is only needed to go back
     to the empirical or real world probabilities. Indeed, it follows from these
arguments  that consideration of this probability amounts to
redefining the evolution equation without changing  the volatility
coefficient
    $ b(t,  x)$   but  replacing the  drift coefficient   to $   a^*(t,  x)=r_tx$,
independent of the initial coefficient
     $  a(t,  x)\eq \mu x$.

     Unfortunately, in general it is not possible to solve the redefined SDE   corresponding
to   the diffusion coefficients  (4) with  $   a^*(t,  x)=r_tx$.
     However,
      it turns out that in the particular case when  $\s^2(t)= 2r(t)$,
i.e., for eq. (8) below, then both this SDE  and the prizing problem can be solved as we next
 show. We shall consider this case and hence we suppose that under the risk neutral
probability  $   \Bbb   P^* $,  $X_t$  evolves via the
 SDE
 $$dX_t=    r_t X_t   dt+     \sqrt{ 2r_t\big(    X_t^2-    c^2\big)   }  d
W_t^*,~~ X_0=x_0>c\eqno (8)$$

  Here $W_t^*$ is a BM with respect to the risk neutral probability.     Technically,
   the existence and nature  of all objects introduced below is a  difficult point.
In the appendix we sketch how  to perform such a
   construction.

  In the sequel all quantities are referred to the
probability $   \Bbb   P^* $ and hence $X_t$ evolves via (8) which
is our fundamental equation.  Further for ease of notation we drop
here and elsewhere the use of $^*$.

The return process $R_t\eq \log X_t/x_0$ is found via It\^{o}'s
rule to satisfy

$$dR_t=     r_t c^2e^{-2R_t}    dt+     \sqrt{2r_t\Big(1- c^2e^{-2R_t}\Big)} d
W_t,~~ R_0=0\eqno (9)$$

Thus only  when  $R_t $ is close to $ 0$ it does behave like a
classical random walk.

 Useful information about the behavior of the price process at $x=c$ follows by   careful
inspection of    the nature of the boundary $x=c$. Consider the
Feller functions
 $\Sigma (c,x), \Omega(c,x)$  defined by
 $$\Sigma (c,x)=\int_c^x    \f{ p(z)}{b^2(z)} dz\int_c^z{dy\over p(y)} ;~
\Omega(c,x)=\int_c^x
   \f{dy}{p(y)}\int_c^y{ p(z) \over b^2(z)}dz.\eqno (10)$$

 where $p(x)\eq   \sqrt{      x^2-    c^2   } $.   The reader is referred for these
matters to   [24]. Notice that the integrand is
 singular since  it has a square root singularity at $x=c$.
 Upon
evaluation of the integrals we find that
 $$\Sigma (c, x) =\Omega(c,x)=\f1{ 4 r
 }  \log^2\big( {x+\sqrt{      x^2-    c^2   }\over c}\big)\eqno (11)$$
 Thus, unlike what happens in the model defined by (3), we have that  $\Sigma (c, x )=  \Omega (c, x)<\infty$  are  finite, corresponding to a  regular
boundary which
  can be both reached  and exited from in finite time with positive probability.

  While Feller
  analysis  shows that the boundary is \it attainable \rm  it does not clarify  if
the process can be continued past the boundary (and hence whether
prices
    below the level $x=c$ can be attained).   Further
  it it is unclear what  is the probability to reach the boundary or how the to
continue the
  process upon hitting the boundary.

  These kind of problems regarding behavior of the process at and past the boundary
  are generically quite difficult to tackle. It turns out that for  our particular
  model the behavior of the process is completely determined.
Actually we have found that the solution to eq. (8) is given in a
fully explicit way   by

 $$  X_t=  c \cosh \big(   \int_0^t \sqrt{2r_s}  dW_s +   \kappa  \Big)\eqno (12)$$

 where $\kappa\eq   \cosh^{-1}  (    x_0/c   )$. A typical path $t\to X_t$ is shown below in Fig. 1.  To prove  (12) let $\tilde X_t\eq
g(Y_t)$ where $$ g(z)=c \cosh \big( z  ), ~Y_t\eq \kappa+\int_0^t
\sqrt{2r_s} dW_s\eqno (13)$$
  Using  It\^{o}'s rule and the fact that $dY_t= \sqrt{2r_t}  dW_t$
 we find that  $\tilde X_t$  has diffusion coefficients   $ \tilde a(t,  x), \tilde
 b(t,  x)$ satisfying at $x=g(z)$
 $$\tilde a(t,  x) = { 2r_t\over 2}\p_{zz} g=r_t g(z)\eq r_tx$$

  $$\tilde b(t,  x) =\sqrt{2r_t} \p_{ z} g= \sqrt{2r_t} c\sinh \big (     z )=
\sqrt{    2r_t\Big(  x^2-    c^2  \Big) } \eqno(14) $$

  i.e. $\tilde X_t\eq g(Y_t)$ solves the SDE (8).

  Notice that the last equality and the fact that the $\sinh$ takes both signs,
imply that the following prescription must be given at the
barrier:
  $$b\big(g(z)\big)=-b\big(g(-z)\big)\eqno(15)$$
  Thus (12) solves (8) provided the square root is defined with   a branch cut  on
$(c,\infty)$.

     We note  that  the conditional density  $f(s,Y|t ,y ), t<s$ of
     $Y_t $ solves
     $$\Big(\p_t + r(t)\p_{yy}\Big)f=0; f( s,y|s,y_0)=\delta
     (y-y_0)\eqno(16)$$
     which is converted into the classical BM or heat equation
     upon time  transformation via $t'=\varphi(t) $ where we define
      $$\varphi(q,r)\eq   2\int_q^r  r_s ds; \varphi(t)\eq \varphi(0,t)\eq 2
\int_0^t  r_s ds\eqno(17)$$

     It follows that the process $Y_t$ has the distribution of a BM evaluated at time
     $t'$ and  hence  we can represent $X_t$ as
  $$  X_t=  c \cosh Y_t \text{ where } Y_t=\kappa+B_{\varphi(t)} \eqno(18)$$

 and   $B_t$ is a new BM. It follows from (12) that     $X_t
\geq  c$ and  that $X_t $ \it attains the barrier $c$ whenever the
process $Y_t$ reaches $0$ \rm, i.e., when $\int_0^t \sqrt{2r_s}
dW_s +\kappa=0$, which, according to (18) happens eventually with
probability one. As pointed out, it follows from (18) that in that
case the process $X_t$ is reflected and hence the level $x=c>0$
acts as a resistance of the stock value.

Let us now obtain $p( T, X|t, x) $, the   probability density
function (pdf) of the price process conditional on the value at
time $t$:  $X_t=x, t<T$. This pdf satisfies the backwards
Kolmogorov-Fokker-Planck equation
$$\Big(\p_t + r(t)(x^2-c^2)\p_{xx}+rx\p_x\Big)p=0;
p(T,X|T,x)=\delta(x-X)\eqno(19)
     $$

     Motivated by  (13,17) above we define new coordinates
     $t'=\varphi(t), y=\log \Big(x+\sqrt{x^2-c^2}\Big)-\log c$. In terms of the new
coordinates $p$ solves
    $$\Big(\p_t + {1\over 2}\p_{yy} \Big)p=0; p(T,Y|T,y)=\delta
  ( c \cosh y-c\cosh Y)   \eqno(20) $$
  Using the well known formula
  $$c\delta
  ( c \cosh y-c\cosh Y)=\Big(\delta
  (   y-  Y)+\delta
  (   y+Y)\Big)/|\sinh Y|$$
  we find that

$$p( T, X|t, x) = {1\over  \sqrt{ 2r_t\big(    X^2-    c^2\big)
}}\un{\pm}\sum\exp\Big(-{1\over 2\varphi(t,T)} \log^2  {
x+\sqrt{x^2-c^2} \over X\pm\sqrt{X^2-c^2}}\Big ) \eqno(21) $$

 We next compare the evolution of prices under this model
and that described by GBM. For a meaningful comparison we need to
have $r_t=r$ constant and,   in Eq. (1), $\mu\eq r, \s=\sqrt{2r}$.
In this case (2) yields
$$X_t^{\text{GBM}}= x_0 e^{\sqrt{2r} W_t}  \eqno(22) $$

 Note that whenever $W_t>>1$ both process behave in a
very similar way:
$$\un{W_t\to \infty}\lim  X_t / X_t^{\text{GBM}}  =  {x_0+\sqrt{x_0^2-c^2} \over  2
x_0}    $$ However as $W_t\to-\infty$ then
$$X_t  \approx   \big(x_0-\sqrt{x_0^2-c^2}\big)/2  e^{  \sqrt 2r|W_t|}\to\infty,
X_t^{\text{GBM}}  =    x_0   e^{  \sqrt 2rW_t}\to  0$$

Finally, we note that  in the limit    $c\to 0$ we can recover (2)
from (18).  A careful calculation shows that
$$\un{c\to 0}\lim X_t=x_0\exp   B_{2 rt} $$
and   hence,   using
 that $\sqrt{2r} W_t$ is a BM at time  $2 rt$,   we recover (2).

 \section{  valuation of   securities}

We    consider   here  the valuation of   securities   earning a
terminal payoff $\Theta_1(X_T)$ at maturity $T$.
       We also allow for the security to pay  dividends at a continuous  rate
$\Theta_2(s,X_s),t\le s\le T$
      where we  suppose  that both    $\Theta_1:  \Bbb R\to \Bbb R$ and $ \Theta_2
:\Bbb R^+\times  \Bbb R\to  \Bbb R$    are continuous.
      The standard case of stock option valuation  corresponds to
      taking  $\Theta_1(x)=(x-k)^+, \Theta_2(x)=0$ where $k$ is
      the strike.

      Let $v_t$ be the  (actual) $t$-price       of    such
    European derivative  maturing at $T$, which  must also  depend    on    $ T$ and the
actual price $ x=X_t$ of the stock.

     We \it assume \rm the
   existence of  risk-neutral probability $\Bbb P^*$
   under which  relative prices of stocks and \it  more generally\rm,   of
self-financing strategies $v'_t\eq  v_t/Z_t$
  are martingales with respect to the history of the process up to time $t$: $
\mathcal F_t$ (notice  that we shall drop again the symbol$^*$).
If this is the case,  reasoning similarly  as in (7) and use of
the martingale property       yields  that

$$      v_t  e^{\int_t^T r_ldl}\eq  Z_T v'_t  = Z_T\Bbb  E
\Big(    v'_T \Big| \mathcal F_t
 \Big)=  \Bbb  E  \Big(   v_T \Big| \mathcal F_t
\Big) $$ To continue further  we note that  $v_t$ must satisfy at
$t=T$  that
     $$v_T =  \Theta_1(X_T) +\int_t^T\Theta_2(s,X_s)Z_T/Z_sds\eq v_T^1+v_T^2
\eqno(25)$$
      as the RHS is precisely the earning  at maturity. Note further that
 $ v_T^1\eq  \Theta_1(X_T)
     $ is obviously Markovian and hence that $ \Bbb  E  \Big(   v_T \Big| \mathcal F_t
\Big) = \Bbb  E  \Big(   v_T \Big|  X_t \Big)$.

     It can be proven that   $ v_T^2 \eq \int_t^T\Theta_2(s,X_s)Z_T/Z_sds
     $   is also a Markov process and hence that it satisfies
     $$\Bbb  E ( \int_t^T \Theta_2(s,X_s)Z_T/Z_s ds \Big|  \mathcal F_t)= \Bbb  E (
\int_t^T \Theta_2(s,X_s)Z_T/Z_s ds \Big|  X_t)\eqno(26)$$

Hence we finally obtain the price of  the security as

$$      v_t  =   \Bbb  E  \Big(    e^{-\int_t^T r_sds} \Theta_1(X_T)
+\int_t^T\Theta_2(s,X_s)Z_t/Z_sds \Big| X_t \Big) \eqno(27)$$ This can be simplified further by  reasoning  as follows. Let $X_s^{ t,x },  t<s$
be the price process at time $s$  knowing that it starts at $x $ at   time $t$, i.e., $ X_s^{ t,x }\eq X_s $ is the solution to (8):

$dX_s=    r_s X_s   ds+     \sqrt{ 2r_s}\big( X_s^2-
c^2\big)^{1/2}      d   W_s , $ with initial condition $X_t=x$.
If we use the well known property ([24])
$$ \Bbb E  \Big(      X_T \Big|X_t=x\Big)= \Bbb  E  \Big( X_{T}^{t,x}\Big)
 \eqno(28)$$
  then      eqs. (13,27) are rewritten   in the convenient form
 $$    X_{T}^{t, x} \eq c \cosh \big(    B_{\varphi(t,T)} + \nu \Big), \nu\eq
\cosh^{-1}(x/c),  $$
$$ \Bbb  E  \Big(     \Theta_1(X_T)\Big|X_t=x\Big)=\Bbb  E  \Big(     \Theta_1\big(
X_{T}^{t, x} \big)\Big)=    \int dY  {\Theta_1\big(c\cosh( Y +\nu)\big)\over \sqrt{2\pi\varphi(t,T)}} e^{-{Y^2\over 2\varphi(t,T)}}\eqno (29)$$
 and so forth (actually, $X_{T}^{t ,X_{t }^{t_0,x_0}} =X_{T}^{t_0,x_0}  $ when $r$
is constant).

 Alternatively, with  (21) at our disposal  we also  obtain that the   price of a
security
  paying    dividends at a continuous  rate   $\Theta_2(s,r_s)$ and  a terminal
value $\Theta_1(r_T)$ at maturity is given in an    explicit
 way by
 $$ v_t=    e^{-\int_t^T r_ldl} \int dX \Theta_1(  X) p( T, X|t,
 x)+ \int_t^Tds\int dX \Theta_2(s,X )p( s, X|t,x)e^{-\int_t^sr_ldl} \eqno(30) $$
 Note that by using the Feynman-Kac theorem    $v_t$   may be also  evaluated by solving
 the backwards equation
$$\Big(\p_t + r(t)(x^2-c^2)\p_{xx}+rx\p_x- r\Big)v=-\Theta_2(t,x)$$
with the terminal condition $\un{t\to T} \lim v_t= \Theta_1(x).$

  As a natural application we evaluate the price of  the plain vanilla
call with strike $k$ corresponding to   $\Theta_1(x)=(x-k)^+,
\Theta_2(x)=0$. Let us introduce
$$\hat x \eq (x +\sqrt{x^2-c^2})/2, ~~\hat  k \eq
(k +\sqrt{k^2-c^2})/2$$
$$N=  {1\over  \sqrt{\varphi(t,T)}} \log{\hat  x\over\hat  k},~~ N_{\pm}\eq N  \pm   \sqrt {
  \varphi(t,T)}, $$
 $$M=  -{1\over  \sqrt{\varphi(t,T)}} \log  {\hat  x \hat k\over c^2 }, M_{\pm}\eq M  \pm \sqrt {
  \varphi(t,T)}\eqno (31)$$
 Then, in terms of  $ \Phi $, the distribution function  of the  normal variable
$\mathcal N(0,1)$,  we
 find from (27) to (30), that, if $X_t=x\ge c$, the plain vanilla
call price is given by
 $$   v_t =
   \hat  x
 \Big(\Phi (  N_+  )+\Phi (   M_- )\Big) +    (x-\hat x)
  \Big(\Phi (  N_-  )+\Phi (   M_+)\Big)  - k  e^{-\int_t^T r(s)ds}    \Big(\Phi (
N   )+\Phi (   M )\Big)   \eqno (32)$$

  The situation when
$c=0$ and $r$ is constant  amounts to having no barrier and hence
(32) must reduce to the BSM formula. Indeed, in this case one has
$ \hat x= x,    \varphi(t,T)=2r(T-t)$,
$$N\eq N^0=  {1\over  \sqrt{ 2r(T-t)}} \log{x \over k },~~ N_{\pm}^0\eq N^0  \pm
\sqrt {    2r(T-t) }, M=-\infty\eqno (33)$$ and
 hence $  \Phi (   M  )=\Phi (   M_{\pm}  )=0$,  most of the terms in  (32)
 drop out and we recover the  BSM formula
$$  v_t   =
   x
 \Phi (  N_+^0  )   - k  e^{- r(T-t)}  \Phi( N^0) \eqno (34)$$

 Another interesting  simplification  appears when the strike
    $k$ coincides with $c$: $k=c$. Using that  for this case
   is $\hat k=c, M=-N,  M_{\pm}=-N_{\mp}$ and the well known property
   $\Phi(z)+\Phi(-z)=1$ we find that all parenthesis in (32) add to $1$ and the price of the  option is   the deterministic    price:

   $v_t=x-c   e^{-\int_t^T r(s)ds} $.

The result is easy to understand; indeed, as we pointed out (12) implies that $X_t  \geq  c$ which rules the possibility to have $k<c$, i.e.,
this case corresponds to having   taken $k$ at its lowest possible value. Thus all uncertainty disappears since  with probability one $X_T>k$ and
the option will be exercised will probability one.

\medskip

 In figure (1) we plot  a typical path of the price process (8) starting at $x_0=5$. We assume a yearly interest rate
   $r=4.5\%$  yr$^{-1}$, annual
volatility $ \s= 30 \%$    and suppose that the support  is    placed at
 $c=4$. Notice how eventually prices get near and  eventually  hit the support level $c=4$ lingering around for some time.

\begin{figure}[!ht]
\begin{center}
\includegraphics[width=12cm,height=7cm]{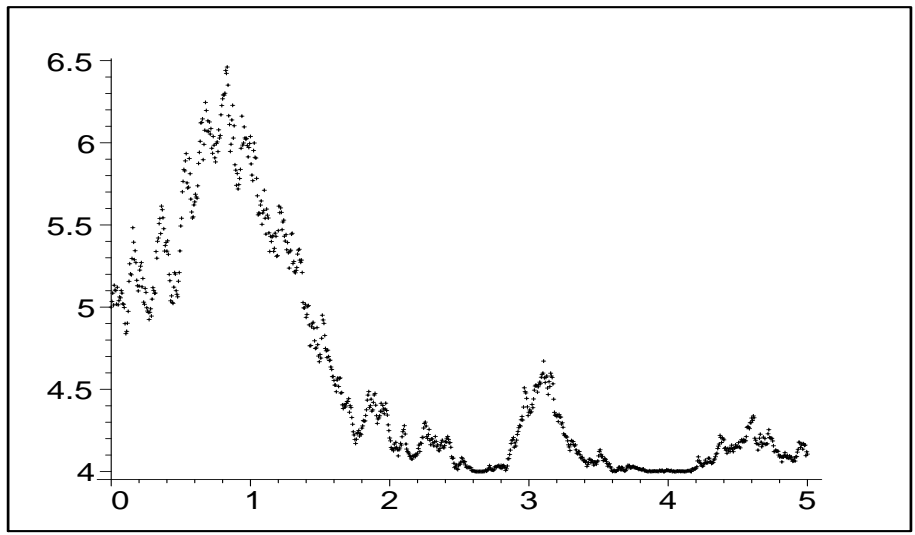}
\caption{    A possible evolution  of the price process. We plot $X_t$ as a  function of time during  a  time span of   five years ($t=5$). The
parameters have been chosen as:  $r=4.5\%$  yr$^{-1}$, $ \s= 30\%$. } \label{path}
\end{center}
\end{figure}

\vskip10.5truecm

In figure   (2)   we plot   the   call price  $v $  in terms of  the initial stock price
 $x $ corresponding  to a constant annual   interest rate $r=4.5\%$ yr $^{-1 } $  with    annual
volatility $ \s= 30 \%$      and
 time to maturity    $ T-t=1 $ year.   The barrier is located at  $c=4$ while the strike  $k=8$. The thick solid
line represents   (32) while the   dotted line is the classical BSM  call price; the deterministic price $d\eq (x- k e^{-\int_t^T r(s)ds})_+$  is
the thin straight line.     In all cases  one finds that the prices  implied by  the classical BSM valuation formula  and (32) are quite similar,
specially  for long $x$. Notice how the   BSM formula always overprices the call option compared with the formula (32). However, it seems that the
variation is only significant in the region $c\le x\le 2c$,  irrespective of how large    $c$ is.

Actually, we find in all cases  that the  deviation  of (32) from  (34) is quite small (see figs.  2 and 3). This is easy to understand
qualitatively when $x$ is long, since then
  $ \hat x  =x-{c^2\over 4x}+O(1/x^{3/2})$ and hence we  find the expansion
  $$ \sqrt{\varphi(t,T)} N=    \log({   x \over  k})+\log \zeta - {c^2\hat k \over 4x}+O(1/x^{3/2}),~~   $$
 where $0\le \zeta \eq \log  k /\hat k \le \log 2$. Further $M\approx (1/\log x)e^{-\log^2 x}$ and
 $$   v_t  \approx
      x - k  e^{-\int_t^T r(s)ds}          +O(1/x )
    $$
    as in the BSM formula.
 However, such a rough argument does not account for the close similarity  even for small and moderate values of x.

The dependence of the option price  upon the  time to maturity  up to  40 years is  shown in figure 3. The plot corresponds to an ATM option for
which moneyness $x/k=1$. The rest of parameters have not been changed.

\begin{figure}[!ht]
\begin{center}
\includegraphics[width=12cm,height=7cm]{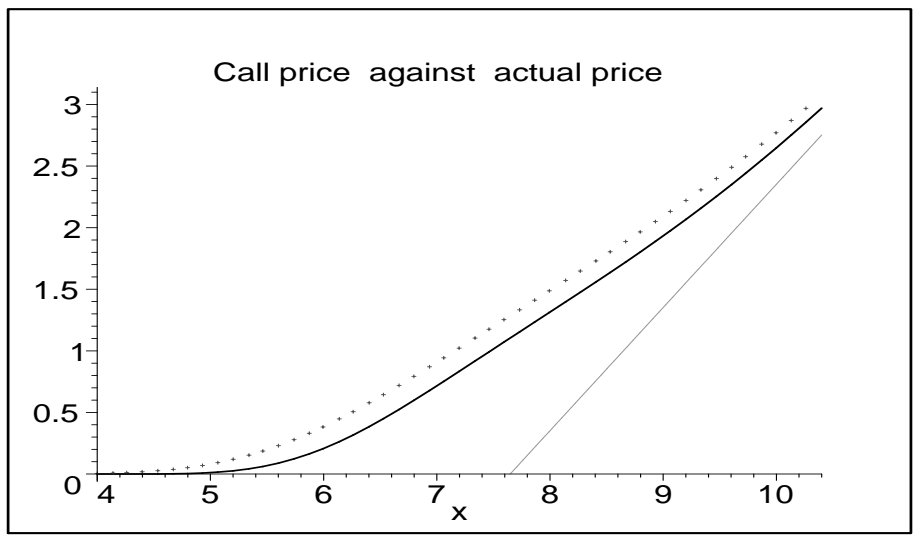}
\caption{   plot  of the    call price  $v $  in terms of      $x $ corresponding to  an annual interest rate  $r=4.5\%$, annual volatility $ \s=
30\%$ and
    $ T-t=1 $year   with  a barrier   at $c=4$. Strike is  taken as $k=8$.
 The thick
line  represents  (32)  while the  dotted line is the  BSM call price. Deterministic price is the thin line} \label{yield-CIR}
\end{center}
\end{figure}

\begin{figure}[!ht]
\begin{center}
\includegraphics[width=12cm,height=7cm]{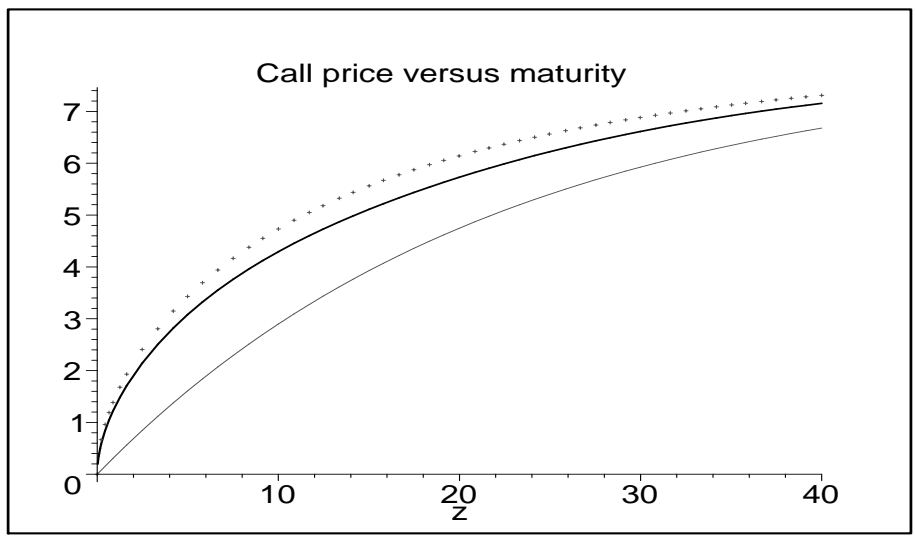}
\caption{   plot  of  an  ATM call option      price  $v $  in terms of   time to maturity $z=T-t$.
 The thick solid
lines represents  (32)  while the   dotted line is the  BSM
 call price. Deterministic price is the thin line. Parameters are as in the previous caption.} \label{yield-CIR}
\end{center}
\end{figure}

\newpage

\section{Appendix}

Consider again eq. (4) $dX_t=     \mu X_t   dt+    b(t,X_t) d W_t $  where $W_t$ is a BM with respect to the real world probability and $b(t,x)=
\sqrt{ 2r_t\big(    X_t^2- c^2\big) }$.
 Let

 $$u(t,x)\eq \Big(r(t) - \mu\Big)
x/b(t,x ), ~    M_t=\exp   \Big(-\int_0^t u(s,   X_s)dW_s- {1\over 2} \int_0^t u^2( s,  X_s)ds\Big) $$
 Here $u(t,x)$ is the so called \it market price of risk \rm. Then if   $X_t$ solves the above  equation
   $M_t$ can be proven formally to be  a  martingale. Note however that
    a rigorous  proof of the latter fact runs into
technical difficulties  due to the singularity of $u$ at $x=c$
which might prevent, in principle, for $M_t$ to be a Martingale.
We  skip a rigorous analysis as we    expect this to be the case.
  Defining the risk neutral
probability $ \Bbb P^*
  $ by   $d   \Bbb   P^*   = M_T d\Bbb  P$
  it follows from  Girsanov's theorem (see [17,20])  that  $
  W_t^* \eq W_t +\int_0^t u (s, X_s)ds $    is a
BM under $ \Bbb    P^* $. In this case an easy calculation shows
that $X_t$ \it also \rm
  satisfies  Eq.
 (8) driven by  $W_t^*   $, a
BM  with respect to $ \Bbb    P^* $.

 \vskip1truecm

 \section{Bibliography}

[1]F. Black, M. Scholes, J. Pol. Econ. {\bf 81} (1973) 637-659.

[2] R.C. Merton, Bell J. Econ. Manage. Sci. {\bf 4} (1973)
141-183.

[3] M.G. Kendall, J. R. Stat. Soc. {\bf 96} (1953) 11-25.

[4] M.F.M. Osborne, Oper. Res. {\bf 7} (1959) 145-173.

  [5]R. N. Mantegna  $\&$ H. E. Stanley,    Nature    {\bf   376}, (1995), 46

[6] B. Mandelbrot, J. Bus. {\bf 35 }(1963) 394

[7]S. Galluzio, G. Caldarelli, M. Marsilli $\&$ Y-C Zhang, \it
Physica A, \rm {\bf 245}, (1997), 423

 [8] A. Matacz, Int. J. Theor. Appl. Finance
{\bf 3 }(2000) 143-160.

[9] J. Masoliver, M. Montero, J.M. Porra,  \it Physica A \rm {\bf
283 }(2000) 559-567.

[10] J. Masoliver, M. Montero, A. McKane, Phys. Rev. E {\bf 64}
(2001) 011110

[11]S. Galluzio, \it Europ. Phys. Jour.  B, \rm  {\bf
20(4)},(2001), 595

  [12]J. Perell\'o $\&$ J. Masoliver, \it Physica A \rm {\bf 314},
(2002), 736

[13]]J. Perell\'o $\&$ J. Masoliver, \it Physica A \rm {\bf 308},
(2002), 420

 [14]J.E. Ingersoll  {\it  Theory of financial decision making}" Rowman $\&$
Littlefield, Savage MD (1987)

 [15]J. Hull, "{\it  Options, futures,  derivatives}",
Prentice Hall Univ.Press, (1997).

 [16]  D. Duffie,  {\it  Dynamic Asset Pricing Theory}",  Princeton University
Press, Princeton (1996)

[17] T. Mikosch,  {\it  Elementary Stochastic Calculus: with
Finance in View}, World Scientific, 1998.

[18] J. Voit, "{\it  The statistical mechanics of financial
markets}", Springer Verlag, Berlin, (2003)

[19]J.Bouchard  $\&$ P.Potters, "{\it  Theory of financial risk,
from statistical physics to risk management}", Cambridge Univ.
Press, Cambridge 2000

[20] G. Vasconzelos,  \it Braz. Jour. Physics \rm {\bf  34(3B)},
(2004), 1039

[21]R. N. Mantegna  $\&$ H. E. Stanley, "{\it  An introduction to
Econophysics}", Cambridge Univ. Press, Cambridge  (2000)

[22]  J.M. Harrison   $\&$ S. Pliska, Stochastic Process. Appl.
{\bf 11} (1981) 215

[23]  J.C. Cox, J.E. Ingersoll  $\&$ S.A.  Ross.   \it
Econometrica \rm {\bf 53 (2)}, (1985) 385.

 [24] W. Horstheeke  and R.   Lefever,   "{\it Noise induced transitions},
 Springer series in synergetics  15,
 Springer Verlag,
 Berlin

\newpage

\end{document}